\documentclass[preprintnumbers,amsmath,amssymb,floatfix,12pt,prd,nofootinbib]{revtex4}

\usepackage{graphicx}
\usepackage{epsfig}
\usepackage{bm}
\usepackage{amsfonts}
\usepackage{subfigure}
\usepackage{multirow}
\usepackage{color}
\usepackage{float}
\usepackage{xfrac}
\usepackage{newtxtext,newtxmath}
\usepackage{xcolor}
\usepackage[T1]{fontenc}
\usepackage{newtxtext,newtxmath}

\begin{document}

\title{\textbf{Trapped sausage modes in magnetic pores: existence criteria and viscous damping}}

\author{Zanyar Ebrahimi}
\email{zebrahimi@maragheh.ac.ir}
\affiliation{Research Institute for Astronomy and Astrophysics of Maragha, University of Maragheh, Maragheh, P.O.Box 55136–553, Iran}

\date{\today}

\begin{abstract}
	The conditions governing the existence and evolution of trapped sausage waves in photospheric magnetic pores are investigated, with particular emphasis on the limitations of analytical cut-off criteria and the role of viscous dissipation. A pore is modelled as a cylindrical magnetic flux tube with smoothly varying radial profiles of magnetic field and density, subject to the constraint that its temperature is lower than that of the surrounding atmosphere. For finite plasma-$\beta$, the linearised equations governing sausage-mode perturbations are derived, and an initial-value problem is solved numerically to obtain the full spatio-temporal evolution of the waves. The applicability of the analytical cut-off relation derived for a flux tube with step-function equilibrium profiles is examined, revealing that its predictions depend on the equilibrium parameters. The simulations reveal distinct wave behaviours across different parameter regimes, including the simultaneous excitation of trapped fast and slow sausage waves for a representative photospheric parameter set. Viscous damping is further investigated through the temporal evolution of the wave energy. Energy dissipation increases with decreasing viscous Reynolds number, increasing longitudinal wavenumber, and increasing radial magnetic-field contrast. These trends are consistent with enhanced viscous damping associated with increasingly fine spatial gradients generated through phase mixing. The findings highlight the importance of radial equilibrium structure in assessing sausage-wave trapping and demonstrate the sensitivity of wave-energy dissipation to the physical parameters of photospheric magnetic pores.
\end{abstract}


\maketitle

\newpage
\section{Introduction}

Recent advancements in high-resolution instrumentation and sophisticated three-dimensional imaging techniques have significantly enhanced our ability to measure and infer key physical parameters of the solar atmosphere \citep[see e.g.][]{Erdelyi2022, Bate2022}. In parallel, the integration of numerical modeling and magnetohydrodynamic (MHD) theory has propelled the development of coronal seismology, offering a powerful diagnostic framework for probing otherwise inaccessible properties of the coronal plasma. Foundational contributions by \cite{Zaitsev1975,Zaitsev1982} and \cite{Edwin1983} laid the theoretical groundwork for MHD wave propagation in magnetic flux tubes, specifying the conditions under which various wave modes can be sustained. Building on these early insights, the study of MHD oscillations in magnetized plasma cylinders has evolved into a robust field, wherein wave-based diagnostics are routinely employed to investigate the plasma structure and magnetic field configuration of the solar atmosphere.

Among the global MHD modes supported by a cylindrical flux tube, the sausage mode (characterized by axial symmetry and azimuthal mode number $m=0$) is of particular interest. This mode manifests as periodic radial expansions and contractions of the tube’s cross-section, accompanied by perturbations in plasma density, velocity, and magnetic field strength. Two distinct types of sausage oscillations can arise: the slow and fast modes \citep{Edwin1983}. For fast sausage waves to remain trapped within the flux tube, their period must lie below a critical threshold known as the cut-off period \citep{Roberts1984}, typically on the order of the Alfvén crossing time which is the characteristic timescale for Alfvén wave propagation across the tube. When the wave period exceeds this cut-off, the flux tube continuously emits fast magnetoacoustic waves into the surrounding medium, leading to a reduction in the amplitude of the fast wave component in the flux tube. In contrast, slow sausage modes remain confined regardless of their period. For a comprehensive review of sausage-mode MHD waves in coronal flux tubes, see \cite{Li2020}.

Observational studies have identified sausage-mode oscillations with periods ranging from 5 to 30 seconds in coronal loops \citep{Nakariakov2003, Melnikov2005, Inglis2008, Srivastava2008, Tian2016}. Similar signatures have also been detected in photospheric structures such as pores and sunspots \citep{Morton2011, Grant2015, Freij2016,Jafarzadeh2024}, indicating the ubiquity of these modes across different layers of the solar atmosphere. Fast sausage waves, in particular, have been proposed as a plausible mechanism underlying the generation of quasi-periodic pulsations with short periods --- on the order of a few seconds --- commonly observed in flare-associated coronal loops \citep[see e.g.][]{Nakariakov2009, Kolotkov2015, Shi2024}.

Extensive theoretical efforts have been devoted to the study of sausage-mode oscillations in magnetic flux tubes, yielding a rich body of literature that explores their dynamics under various physical conditions \citep[see e.g.][among many others]{Erdelyi2007, Pascoe2007, Inglis2009, Nakariakov2012, Pardi2013, Farahani2014, Li2014, Oliver2015, Bahari2017, Bahari2018, Sadeghi2019, Lopin2019, Guo2021,Li2022,Wang2023}. In the zero-$\beta$ limit --- where plasma pressure is neglected in favor of magnetic pressure --- \cite{Nakariakov2012} demonstrated that in cylindrical geometry the wave period becomes independent of the wavenumber in the long-wavelength regime. Instead, the period is governed by parameters characterizing radial plasma inhomogeneity, such as the steepness of the density gradient and the contrast between internal and external plasma densities. Moreover, plasma-$\beta$ has been shown to exert a negligible influence on the cut-off wavenumber \citep{Inglis2009,Chen2016}.

In addition to the well-known mechanism of wave energy leakage associated with fast sausage modes, resonant absorption of slow sausage waves can also play a significant role in the damping process. While fast modes typically lose energy through radiation into the external medium, slow modes may undergo resonant coupling with localized slow continua in inhomogeneous plasma environments \citep[see e.g.][]{Giagkiozis2016, Yu2017, Chen2018, Sadeghi2019, Yu2024}. The investigation of resonant absorption of axisymmetric modes ($m=0$) within the Alfv\'en continuum, in the presence of a twisted magnetic field, revealed that both the longitudinal magnetic field and the plasma density play critical roles in wave damping \citep{Giagkiozis2016}. In the absence of an azimuthal magnetic field component, resonant absorption for axisymmetric modes can occur only through the slow (cusp) resonance. This coupling facilitates the transfer of wave energy from global mode to localized phase-mixed oscillations in the inhomogeneous region of the flux tube \citep{Ebrahimi2024}, which may ultimately result to efficient energy dissipation due to dissipative effects like viscosity or magnetic diffusivity. Resonant absorption of slow sausage waves in the slow continuum has been investigated by \cite{Yu2017} in thin inhomogeneous layer (thin boundary) approximation. They showed that resonant damping of the slow surface sausage mode within the slow continuum --- previously underestimated in earlier interpretations --- can be efficient under magnetic pore conditions. \cite{Chen2018} investigated the influence of resistivity and resonant coupling to the cusp continuum on the damping of linear slow surface sausage modes in a photospheric cylindrical flux tube. Their results indicated that, in general, damping due solely to the cusp resonance is relatively weak, with the minimum damping time to period ratio reaching values on the order of 100. They showed that, resistivity can be an order of magnitude more effective, making it an essential factor in the analysis of the damping of slow surface sausage modes.

There has been increasing interest in recent years in sausage-mode oscillations within photospheric waveguides, such as magnetic pores and sunspots, particularly concerning the spatial damping of propagating sausage waves and their potential role in energy transport and dissipation. Despite this attention, numerical studies that incorporate both viscosity and a finite plasma-$\beta$ in radially non-uniform flux tubes remain limited. Such investigations can help clarify the role of dissipative processes in the dynamics and energy evolution of MHD waves in radially structured photospheric waveguides.

Solar pores are small magnetic features, typically $1-6~Mm$ in diameter, with field strengths on the order of a kilo gauss \citep{Sobotka2003}. Similar to sunspots, pores appear darker than the quiet Sun, with mean intensities measured up to 40\% below the surrounding photosphere \citep{Verma2014}. Observations indicate that pores form when the local intensity falls below 0.85 of the ambient photospheric level and the magnetic field strengthens to about 650 G \citep{Dorotovic2014}. Unlike sunspots, pores lack penumbrae, representing relatively simple magnetic structures with lifetimes ranging from several hours to a few days \citep{Sütterlin1996}.  These structures have traditionally been regarded and frequently modeled as simple, isolated magnetic flux tubes \citep{Simon1970, Hurlburt2000,Sadeghi2019}. There is evidence of sausage waves in pores, as reported in several observational studies \citep{Morton2011,Grant2015, Keys2018, Grant2022,Jafarzadeh2024}. This motivate us to study the propagation of sausage waves in photospheric pores.

The present work extends the analysis of \cite{Ebrahimi2024} with a specific focus on the trapped sausage modes in magnetic pores and introducing viscosity into the modeling framework.  The problem formulation is outlined in Section \ref{model}, followed by the presentation of numerical results in Section \ref{results}. Section \ref{conclusion} provides a summary of the study and concludes with final remarks.


\section{Model and Equations of Motion}\label{model}

To model a magnetic flux tube in its simplest form, we consider a straight cylindrical structure with a circular cross section. The system is described using cylindrical coordinates $(r, \varphi, z)$ where the 
$z$-axis is aligned with the axis of the flux tube. A radially non-uniform background magnetic field is assumed to be oriented along the $z$-direction and is given by $\mathbf{B_0}=B_0(r) \hat{z}$, where
\begin{equation}
	B_0(r)=B_{\infty}+(B_{00}-B_{\infty})\exp[-\left(\frac{r}{R}\right)^\epsilon],
\end{equation}
Here, $B_{00}$ and $B_{\infty}$ denote the magnetic-field strengths on the tube axis and far from the tube, respectively. The parameter $\epsilon>0$ controls the steepness of the radial magnetic field profile around the tube radius $r=R$.

The equilibrium plasma density is likewise assumed to vary radially according to
\begin{equation}
	\rho_0(r)=\rho_{\infty}+(\rho_{00}-\rho_{\infty})\exp[-\left(\frac{r}{R}\right)^\alpha],
\end{equation}
where $\rho_{00}$ and $\rho_{\infty}$ denote the plasma densities at the tube axis and far from the tube, respectively. Here, $\alpha>0$ controls the steepness of the radial density gradient near $r=R$. We define the density contrast as $\zeta\equiv\rho_{00}/\rho_{\infty}$. While the value of $\zeta$ typically ranges between 2 and 20 in over-dense coronal flux tubes \citep[see e.g.][]{Aschwanden2003}, it falls below 1 in under-dense photospheric magnetic structures \citep{Grant2015}.

\begin{figure}
	\includegraphics[width=0.5\columnwidth]{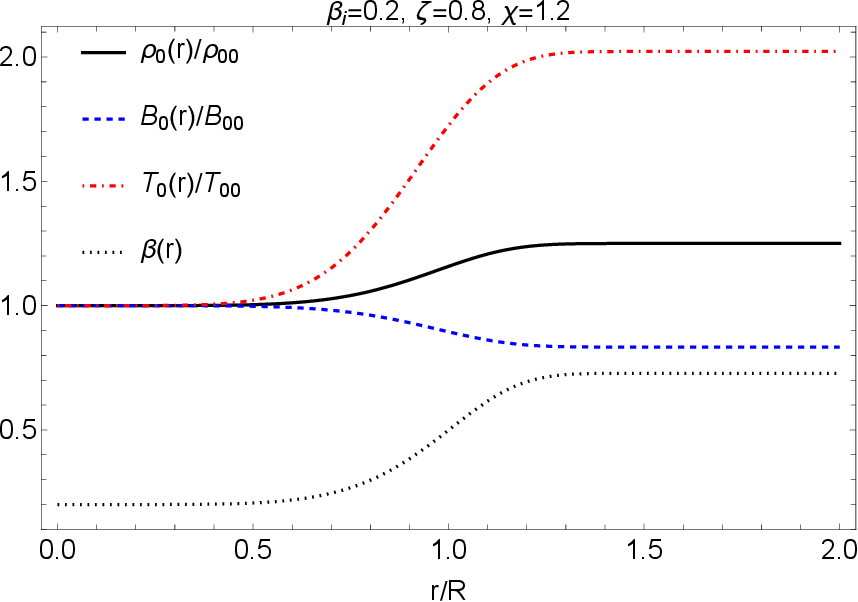}
	\caption{\linespread{1}\selectfont Normalized radial profiles of plasma density (black solid line), magnetic field (blue dashed line), and temperature (red dot-dashed line) as well as the plasma-$\beta$ (black dotted line) for $\beta_i=0.2$, $\zeta = 0.8$, $\chi=1.2$ and $\alpha=\epsilon = 6$.}
	\label{fig:model}
\end{figure}

From the condition of magnetohydrostatic equilibrium, namely
\begin{equation}\label{pb}
	\frac{d}{d r}\left(P_0(r)+\frac{B_0^2(r)}{2\mu_0}\right)=0,
\end{equation}
the profile of gas pressure is obtained as
\begin{equation}\label{equi}
	P_0(r)=P_{00}+\frac{B_{00}^2-B_0^2(r)}{2\mu_0}.
\end{equation}
The equilibrium temperature profile is then determined from the ideal-gas law,
\begin{equation}
	T_0(r)=\frac{P_0(r)}{R_s\rho_0(r)},
\end{equation}
where $R_s$ is the specific gas constant. It is convenient to characterize the equilibrium by the plasma beta on the pore axis,
\begin{equation}\label{betai}
	\beta_i\equiv\frac{2\mu_0 P_{00}}{B_{00}^2},
\end{equation}
and the magnetic-field contrast $\chi\equiv B_{00}/B_{\infty}$. In the solar photosphere, the plasma-$\beta$ typically exceeds unity, with values ranging from about 0.2 up to 100, indicating that gas pressure dominates over magnetic pressure in most regions. \citep{Gary2001,aschwanden2005}. Since we need the pore to be cooler than its environment it is appropriate to obtain the internal to external temperature ratio in terms of the model parameters as
\begin{equation}
	\frac{T_{00}}{T_{\infty}}
	=\frac{1}{\zeta} \frac{\beta_i}
	{1+\beta_i-\chi^{-2}},
\end{equation}
Hence, the condition $T_{00}<T_{\infty}$ for an underdense pore ($\zeta<1$) with an enhanced magnetic field ($\chi>1$)  places a constraint on parameters $\zeta$, $\chi$, and $\beta_i$ as
\begin{equation}\label{betamax}
	\beta_i<\beta_{max}\equiv\frac{\zeta(\chi^2-1)}{\chi^2(1-\zeta)}.
\end{equation}
As an instance, for $\zeta=0.8$ and $\chi=1.2$ we have $\beta_{\max}\simeq 1.22$. Hence, the condition $T_{00}<T_{\infty}$ requires $\beta_i<\beta_{\max}\simeq 1.22$  for $\zeta=0.8$ and $\chi=1.2$. Note that, in the limit $\chi\rightarrow\infty$ we get $\beta_{max}=\zeta/(1-\zeta)$. Thus, for pores with a high magnetic field contrast, \(\beta_{\max}\) depends predominantly on the density contrast, with only a weak dependence on the magnetic field contrast. Fig. \ref{fig:model} illustrates the radial dependence of the plasma density, magnetic field, temperature and plasma-$\beta\equiv 2\mu_0 P_0(r)/B_0^2(r)$ for $\alpha=\epsilon=6$, $\beta_i=0.2$, $\zeta=0.8$ and $\chi=1.2$. It is clear from Fig. 1 that the parameter $\beta(r)$ increases with radius. Using Eqs. \eqref{equi} and \eqref{betai} one can obtain the exterior value of plasma-$\beta$ as follows
\begin{equation}\label{beta}
	\beta_\infty=\chi^2(1+\beta_i)-1.
\end{equation}
Therefore, for a given value of $\beta_i$ satisfying the constraint \eqref{betamax}, an increase in the magnetic-field contrast leads to an increase in the exterior plasma-$\beta$.

The linearized MHD equations of motion for a non-zero $\beta$ plasma in presence of viscosity are as follows
\begin{equation}\label{mhd1}
	\frac{\partial \rho_1}{\partial t}+\mathbf{v}_1\cdot\nabla\rho_0+\rho_0\nabla\cdot\mathbf{v}_1=0,
\end{equation}
\begin{equation}\label{mhd2}
	\begin{split}
		\rho_0\frac{\partial\mathbf{v}_1}{\partial t}=&-\nabla p_1+\frac{1}{\mu_0}\left[\left(\nabla \times\mathbf{B}_0\right)\times\mathbf{B}_1+\left(\nabla \times\mathbf{B}_1\right)\times\mathbf{B}_0\right]\\
		&+\rho_0\nu\left[\nabla^2\mathbf{v}_1+\frac{1}{3}\nabla(\nabla\cdot\mathbf{v}_1)\right],
	\end{split}
\end{equation}
\begin{equation}\label{mhd3}
	\frac{\partial \mathbf{B}_1}{\partial t}=\nabla\times\left(\mathbf{v}_1\times\mathbf{B}_0\right),
\end{equation}
\begin{equation}\label{mhd4}
	\frac{\partial p_1}{\partial t}+\mathbf{v}_1\cdot\nabla P_0-\frac{\gamma P_0}{\rho_0}\left(\frac{\partial \rho_1}{\partial t}+\mathbf{v}_1\cdot\nabla\rho_0\right)=0,
\end{equation}
where, $\mathbf{v}_1$, $\mathbf{B}_1$, $\rho_1$, and $p_1$ denote the Eulerian perturbations of velocity, magnetic field, plasma density, and gas pressure, respectively. Here, $\gamma (= 5/3)$ and $\nu$ represent  the adiabatic index, and the kinematic viscosity. Note that the viscous-heating term is neglected in the linearized energy equation \eqref{mhd4} because, in the absence of a background flow ($\mathbf{v}_0=0$), its leading non-zero contribution is of second order in the perturbations. In addition to viscous dissipation, other non-ideal processes, such as magnetic resistivity and ion-neutral interactions, may contribute to the damping of MHD waves in the partially ionized photosphere \citep[see e.g.][]{Khomenko2016,Stepanov2022}. These processes are not considered here, as the present study focuses specifically on the contribution of viscosity to the damping associated with phase mixing. For mathematical simplicity, an isotropic kinematic viscosity is adopted as an effective description of viscous dissipation. This approximation does not account for the anisotropy of transport coefficients expected in a strongly magnetized plasma \citep[see][]{Hollweg1986}, while allowing the viscous contribution to the wave damping to be isolated without introducing the additional complexity associated with a fully anisotropic transport model.

For axisymmetric sausage modes, the velocity and magnetic field perturbations are confined to the radial and longitudinal directions, such that
\begin{eqnarray}
	\mathbf{v_1} = v_r\,\hat{r} + v_z\,\hat{z}, \label{vpert} \\
	\mathbf{B_1} = b_r\,\hat{r} + b_z\,\hat{z}, \label{bpert}
\end{eqnarray}
where $\hat{r}$ and $\hat{z}$ denote the unit vectors in the radial and longitudinal directions, respectively.
After eliminating $\rho_1$ by combining Eqs. \eqref{mhd1} and \eqref{mhd4}, the resulting set of equations for $v_r$, $v_z$, $b_r$, $b_z$, and $p_1$ is given by
\begin{equation}\label{mhdvr}
	\begin{split}
		\rho_0&\frac{\partial v_r}{\partial t}=-\frac{\partial p_1}{\partial r}-\frac{B_0'}{\mu_0}b_z		+\frac{B_0}{\mu_0}\left(\frac{\partial b_r}{\partial z}-\frac{\partial b_z}{\partial r}	\right)	\\	&+\rho_0\nu\left[\frac{4}{3}\left(\frac{\partial^2 v_r}{\partial r^2}+\frac{1}{r}\frac{\partial v_r}{\partial r}-\frac{v_r}{r^2}\right)+\frac{1}{3}\frac{\partial^2 v_z}{\partial r \partial z}+\frac{\partial^2 v_r}{\partial z^2}\right],
	\end{split}
\end{equation}
\begin{equation}\label{mhdvz}
	\begin{split}
		\rho_0&\frac{\partial v_z}{\partial t}=-\frac{\partial p_1}{\partial z}-\frac{B_0'}{\mu_0}b_r\\
		&+\rho_0\nu\left[\frac{\partial^2 v_z}{\partial r^2}+\frac{1}{r}\frac{\partial v_z}{\partial r}+\frac{4}{3}\frac{\partial^2 v_z}{\partial z^2}+\frac{1}{3}\frac{\partial^2 v_r}{\partial r \partial z}+\frac{1}{3}\frac{v_r}{r}\right],
	\end{split}
\end{equation}
\begin{equation}\label{mhdbr}
	\frac{\partial b_r}{\partial t}=B_0\frac{\partial v_r}{\partial z},
\end{equation}
\begin{equation}\label{mhdbz}
	\frac{\partial b_z}{\partial t}=-\frac{1}{r}\frac{\partial}{\partial r}\left(rB_0v_r\right),
\end{equation}
\begin{equation}\label{mhdp1}
	\frac{\partial p_1}{\partial t}=-v_r P_0'-\gamma P_0\left[\frac{1}{r}\frac{\partial(rv_r)}{\partial r}+\frac{\partial v_z}{\partial z}\right],
\end{equation}
where, the prime sign denotes a derivative with respect to $r$.

Since the equilibrium quantities are uniform along the $z$-direction, we perform a Fourier decomposition in this coordinate and express the $z$-dependence of the perturbations as
\begin{equation}\label{fourier}
	\begin{split}
		&v_r(r,z,t)=\hat{v}_r(r,t)\sin(k_z z),\\ 
		&v_z(r,z,t)=\hat{v}_z(r,t)\cos(k_z z),\\
		&b_r(r,z,t)=\hat{b}_r(r,t)\cos(k_z z),\\
		&b_z(r,z,t)=\hat{b}_z(r,t)\sin(k_z z),\\
		&p_1(r,z,t)=\hat{p}_1(r,t)\sin(k_z z),\\
	\end{split}
\end{equation}
where, $k_z$ denotes the longitudinal wavenumber. Using the Fourier representation given by Eq. \eqref{fourier} and introducing the appropriate dimensionless variables, Eqs. \eqref{mhdvr}-\eqref{mhdp1} can be recast in the following form
\begin{equation}\label{mhdvr2}
	\begin{split}
		\rho_0&\frac{\partial \hat{v}_r}{\partial t}=-\frac{1}{2}\frac{\partial \hat{p}_1}{\partial r}-B_0' \hat{b}_z	-B_0\left(k_z \hat{b}_r+\frac{\partial \hat{b}_z}{\partial r}\right)\\	
		&+\frac{\rho_0}{R_v}\left[\frac{4}{3}\left(\frac{\partial^2 \hat{v}_r}{\partial r^2}+\frac{1}{r}\frac{\partial \hat{v}_r}{\partial r}-\frac{\hat{v}_r}{r^2}\right)-\frac{1}{3}k_z\frac{\partial \hat{v}_z}{\partial r}-k_z^2 \hat{v}_r\right],
	\end{split}
\end{equation}
\begin{equation}\label{mhdvz2}
	\begin{split}
		\rho_0&\frac{\partial \hat{v}_z}{\partial t}=-\frac{1}{2}k_z \hat{p}_1-B_0' \hat{b}_r\\
		&+\frac{\rho_0}{R_v}\left[\frac{\partial^2 \hat{v}_z}{\partial r^2}+\frac{1}{r}\frac{\partial \hat{v}_z}{\partial r}-\frac{4}{3} k_z^2 \hat{v}_z+\frac{1}{3}k_z\frac{\partial \hat{v}_r}{\partial r}+\frac{1}{3}\frac{\hat{v}_r}{r}\right],
	\end{split}
\end{equation}
\begin{equation}\label{mhdbr2}
	\frac{\partial \hat{b}_r}{\partial t}=B_0 k_z \hat{v}_r,
\end{equation}
\begin{equation}\label{mhdbz2}
	\frac{\partial \hat{b}_z}{\partial t}=-\frac{1}{r}\frac{\partial}{\partial r}\left(rB_0 \hat{v}_r\right),
\end{equation}
\begin{equation}\label{mhdp12}
	\frac{\partial \hat{p}_1}{\partial t}=2 B_0 B_0' \hat{v}_r-\gamma (1+\beta_i-B_0^2)\left[\frac{1}{r}\frac{\partial(r \hat{v}_r)}{\partial r}-k_z \hat{v}_z\right].
\end{equation}
where $R_v\equiv C_{A0} R/\nu$ is the Reynolds number, with $C_{A0}=B_{00}/\sqrt{\mu_0\rho_{00}}$ denoting the Alfv\'{e}n speed on the tube axis.  In Eqs. \eqref{mhdbr2}-\eqref{mhdp12}, velocities, lengths, time, magnetic-field strengths, and plasma density are normalized by $C_{A0}$, $R$, Alfv\'{e}n crossing time $\tau_A\equiv R/C_{A0}$, $B_{00}$ and $\rho_{00}$, respectively. Note that, in Eq.~\eqref{mhdp12}, we have used the dimensionless relations $P_0=1+\beta_i-B_0^2$ and $P_0'=-2B_0B_0'$, obtained from Eqs. \eqref{pb} and \eqref{equi}, to eliminate the equilibrium pressure in favor of the background magnetic field.

\section{Solution and Results}\label{results}
We numerically solve an initial value problem for the system of equations (\ref{mhdvr2})-(\ref{mhdp12}), prescribing initial perturbations appropriate for sausage wave as
\begin{equation} \label{init}
	\begin{split}
		&\hat{v}_r(r,0)=r \exp (-r^2/R^2), \hat{v}_z(r,0)=0,\\
		&\hat{b}_r(r,0) =0,\hat{b}_z(r,0)=0,\\
		&\hat{p}_1(r,0)=0.
	\end{split}
\end{equation}
where the initial radial velocity profile is chosen to ensure that the loop axis remains undisturbed. Furthermore, the perturbation amplitudes are set to zero at $r=\infty$, leading to the following kinematic boundary conditions
\begin{equation} 
	\begin{split}
		&\hat{v}_r(0, t)=\hat{b}_r(0,t)=0,\\
		&\hat{v}_r(\infty,t)=\hat{v}_z(\infty,t)=\hat{b}_r(\infty,t)=\hat{b}_z(\infty,t)=\hat{p}_1(\infty,t)=0,
	\end{split}
\end{equation}

Since the conditions at $r=\infty$ is impractical for numerical implementation, we replace it with a finite outer boundary at $r=D\gg R$. Choosing a sufficiently large value for $D$ allows the simulation to proceed for an extended duration before the perturbations reach the boundary and begin to reflect due to wave leakage. For the purposes of this study, we set $D=100R$.

The simulations are performed within a computational domain defined by 
$0<r<100R$ and $0<t<t_{max}$ , utilizing the PDE interface of COMSOL Multiphysics\textregistered (version 6.2). COMSOL employs the finite element method to solve systems of coupled partial differential equations, with adaptive mesh refinement capabilities that enhance resolution where needed. To accurately resolve the steep radial gradients induced by phase mixing, we additionally apply mesh refinement in the region $0<r<2R$.
\begin{figure}
	\includegraphics[width=0.5\columnwidth]{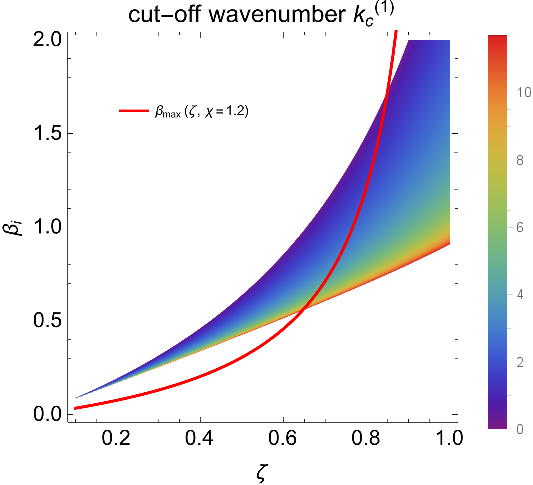}
	\caption{\linespread{1}\selectfont Cut-off wavenumber of the fundamental radial mode of fast sausage wave in the $\zeta-\beta_i$ plane for $\chi=1.2$. Coloured and white regions denote real and complex values of the cut-off wavenumber, respectively. The red curve denotes $\beta_{\max}$ versus $\zeta$.}
	\label{fig:kc}
\end{figure}

Two types of sausage waves can be excited in a flux tube: i) Fast modes, characterized by dominant radial perturbations. ii) Slow modes, in which axial perturbations are dominant. The slow modes are always trapped within the flux tube, whereas fast modes with wavenumbers below a certain cut-off value leak out of the tube. For a flux tube with step-function radial profiles of plasma density and magnetic filed, i.e. $\alpha=\epsilon=\infty$, the cut-off wavenumber is given by \citep{Roberts1984}:
\begin{equation}
	k_c^{(n)}=\frac{\eta_n}{R}\left[\frac{\left(C_{s,i}^2+C_{A,i}^2\right)\left(C_{A,e}^2-C_{t,i}^2\right)}{\left(C_{A,e}^2-C_{A,i}^2\right)\left(C_{A,e}^2-C_{s,i}^2\right)}\right]^{1/2},\label{kc}
\end{equation}
where
\begin{align}
	&C_s(r)=\sqrt{\gamma P_0/\rho_0},\\
	&C_A(r)=B_0/\sqrt{\mu_0\rho_0},\\
	&C_t(r)=\sqrt{C_s^2 C_A^2/(C_s^2 + C_A^2)},
\end{align}
are the sound, Alfv\'{e}n, and tube (slow) speeds, respectively. The subscripts $i$ and $e$ denote quantities inside and outside the flux tube, respectively and $n$ denotes the radial mode number. The expression \eqref{kc} provides the cut-off wavenumber for the $n$th radial mode of the sausage wave, which has $n-1$ zeros in the interval $(0,R)$. The parameter $\eta_n$ is the $n$th root of the Bessel function $J_0(\eta)$, with $\eta_n=2.40, 5.52, 8.65,\ldots$ for $n=1,2,3,\ldots$, respectively. Thus, higher radial modes are associated with larger cut-off wavenumbers. For the fundamental radial mode ($n=1$), the cut-off wavenumber expressed in terms of $\gamma$, $\beta_i$, $\chi$ and $\zeta$ becomes
\begin{equation}
	k_c^{(1)}=2.4\left[\frac{\zeta\left(1+\tfrac{1}{2}\gamma\beta_i\right)-\tfrac{1}{2}\gamma\beta_i\chi^2}{\left(\zeta-\chi^2\right)\left(\frac{\zeta}{\chi^2}-\tfrac{1}{2}\gamma\beta_i\right)}\right]^{1/2}.\label{kc1}
\end{equation}
For example, for $\chi=1.2$, $\zeta=0.8$ and $\beta_i=1$ we get $k_c^{(1)}=2.9$. In practice, for a fixed longitudinal wavenumber $k_z$, radial modes satisfying $k_c^{(n)}<k_z$ remain confined within the system, whereas those with $k_c^{(n)}>k_z$ are radiative and leak out. The initial perturbation of $\hat{v}_r$ in Eq. \eqref{init} is chosen to primarily excite the fundamental radial mode. Nevertheless, the excitation of higher-order radial harmonics due to this perturbation is inevitable. These harmonics, however, possess amplitudes much smaller than that of the fundamental oscillation and dissipate from the loop quickly \citep{Terradas2005, Inglis2009}.

Figure \ref{fig:kc} displays the cut-off wavenumber $k_c$ of the fundamental radial mode across the $\zeta-\beta_i$ parameter space, assuming $\chi=1.2$. The coloured region corresponds to real values of $k_c$, for which $C_{t,i}^2 < C_{A,e}^2 < C_{s,i}^2$ (see Eq. \ref{kc}). Note that, because the present model assumes $\zeta<1$ and $\chi>1$, we always have $C_{A,e}^2<C_{A,i}^2$. The figure also contains two distinct white regions, corresponding to imaginary values of $k_c$. In the upper white region, the characteristic speeds satisfy $C_{A,e}^2 < C_{t,i}^2 < C_{s,i}^2$ whereas in the lower white region, $C_{A,e}^2 > C_{s,i}^2 > C_{t,i}^2$. In the Figure, the red curve indicates the maximum allowable value of $\beta_i$, as obtained from Eq. \eqref{betamax} under the constraint $T_{00}<T_\infty$.

The two white regions in Fig. \ref{fig:kc} require some clarification. The cut-off wavenumber derived by \cite{Roberts1984} is obtained by imposing the limiting condition at which the external radial solution changes from evanescent to propagating. For the fundamental sausage mode, this condition leads to the cut-off relation given in Eq. \eqref{kc}. A real cut-off wavenumber is obtained only when the quantity under the square root is positive. In the upper white region, where $C_{A,e}^2<C_{t,i}^2<C_{s,i}^2$ the quantity under the square root is negative, and the resulting $k_c$ is therefore imaginary. This imaginary value should not be interpreted as a physical cut-off wavenumber. Rather, it indicates that the cut-off condition used in the \citep{Roberts1984} derivation cannot be satisfied by a real longitudinal wavenumber. Consequently, Eq. \eqref{kc} does not provide a physical criterion for the trapped-to-leaky transition in this parameter regime. The same conclusion applies to the lower white region, where $C_{A,e}^2>C_{s,i}^2>C_{t,i}^2$ and expression \eqref{kc} again yields an imaginary $k_c$.

\subsection{Ideal MHD}
Here, we study the propagation of the sausage mode in the absence of viscosity. Since the behaviour of sausage waves depends strongly on the equilibrium model and its parameters, we consider three different parameter sets, $(\beta_i,\zeta,\chi)=(0.2,0.8,1.2)$, $(1,0.8,1.2)$, and $(0.3,0.567,3.012)$, to investigate the wave behaviour under different equilibrium conditions. The first two parameter sets are chosen to examine the effect of the plasma beta at the tube axis, $\beta_i$, while keeping $\zeta$ and $\chi$ fixed. The third set follows the assumptions of \cite{Edwin1983} (see also \citealp{Chen2018}), which provide typical values for photospheric magnetic flux tubes as $C_{s,i}=0.5 C_{A,i}$, $C_{s,e}=0.75 C_{A,i}$, and $C_{A,e}=0.25 C_{A,i}$. For all three sets, we assume $\alpha=\epsilon=6$. According to Eq. \eqref{betamax}, the maximum permissible value of the plasma beta at the tube axis is $\beta_{\max}=1.22$ for $(\zeta,\chi)=(0.8,1.2)$ and $\beta_{\max}=1.165$ for $(\zeta,\chi)=(0.567,3.012)$. In the step-function limit ($\alpha=\epsilon\rightarrow\infty$), Eq. \eqref{kc1} yields a real cut-off wavenumber, $k_c^{(1)}=2.9$, for $(\beta_i,\zeta,\chi)=(1,0.8,1.2)$, whereas it yields imaginary values of $k_c^{(1)}$ for the other two parameter sets, $(0.2,0.8,1.2)$ and $(0.3,0.567,3.012)$. Throughout the remainder of this paper, times, locations, wavenumbers, and frequencies are expressed in units of $\tau_A$, $R$, $R^{-1}$, and $\tau_A^{-1}$, respectively.
\begin{figure*}
	\begin{tabular}{ll}
		(a) & (b)\\
		\includegraphics[width=0.4\columnwidth]{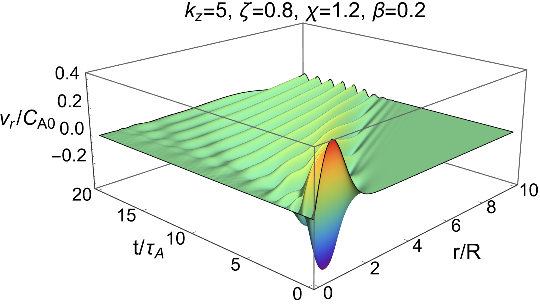}&\includegraphics[width=0.4\columnwidth]{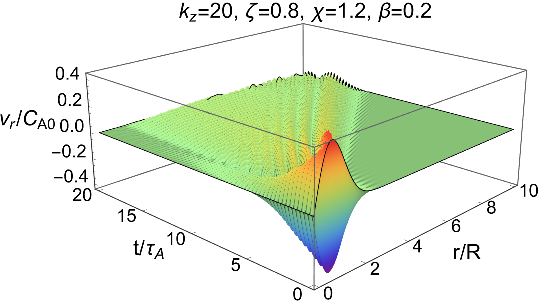}\\
		(c) & (d)\\
		\includegraphics[width=0.4\columnwidth]{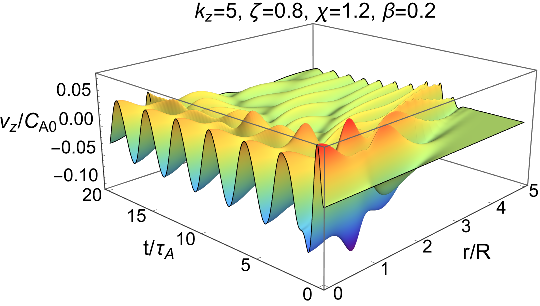}&\includegraphics[width=0.4\columnwidth]{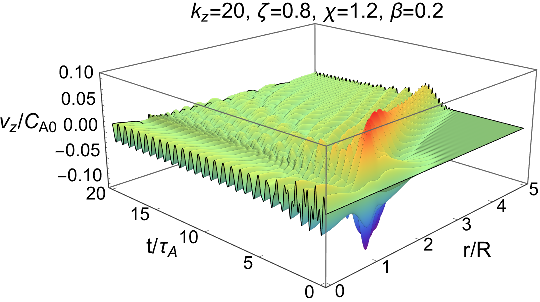}\\
		(e) & (f)\\
		\includegraphics[width=0.4\columnwidth]{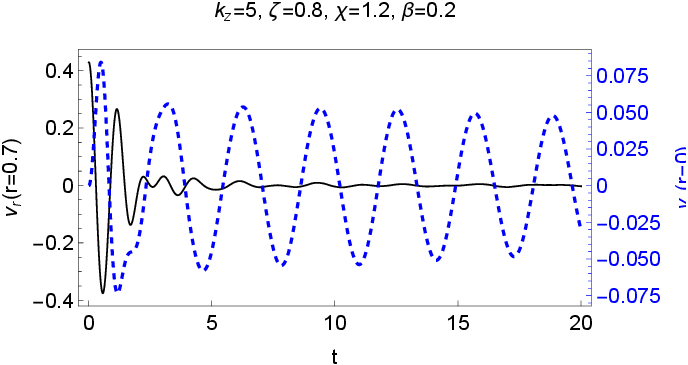}&\includegraphics[width=0.4\columnwidth]{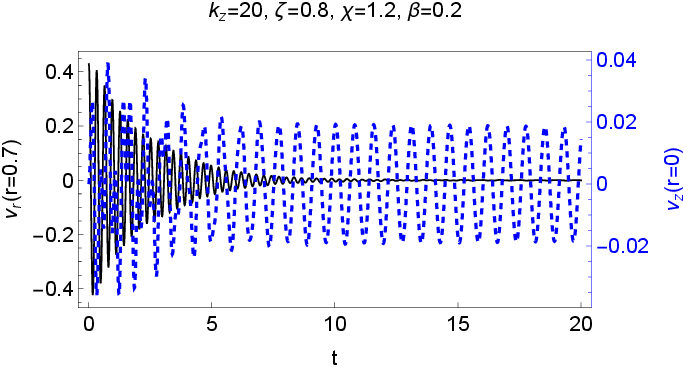}\\
		(g) & (h)\\
		\includegraphics[width=0.4\columnwidth]{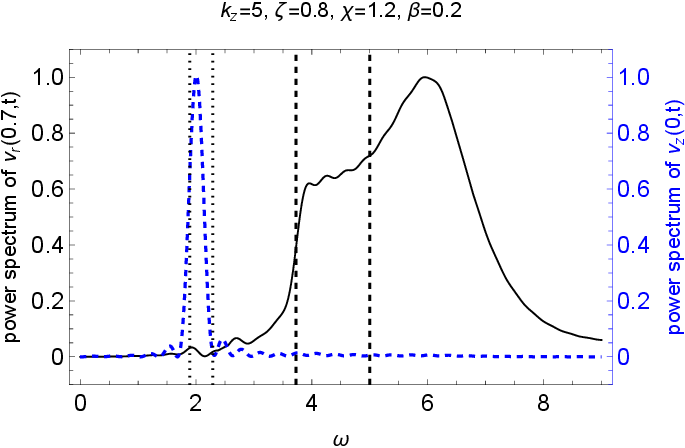}&\includegraphics[width=0.4\columnwidth]{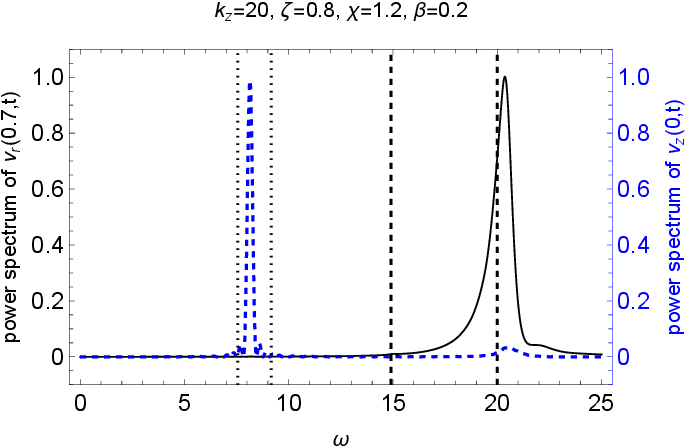}
	\end{tabular}
	\caption{\linespread{1}\selectfont Evolution of the sausage wave for $\zeta=0.8$, $\chi=1.2$, $\beta_i=0.2$ and two values of wavenumber $k_z=5$ (left panels) and $k_z=20$ (right panels). First row (from top): 3D view of the radial velocity perturbation in $r-t$ plane; Second row: 3D view of the axial velocity perturbation in $r-t$ plane; Third row: temporal behaviour of radial velocity perturbation at $r=0.7$ (solid line) and longitudinal one at $r=0$ (dashed line); Fourth row: Power spectra of $v_r(0.7,t)$ (solid line) and $v_z(0,t)$ (dashed line). In panels (g) and (h), the powers are normalized to their maximum values; the vertical dashed and dotted lines denote the intervals of the Alfv\'{e}n and slow frequencies from $r=0$ to $r=\infty$, respectively. Other auxiliary parameters are the same as in Fig. \ref{fig:model}. \normalsize}
	\label{fig:beta02}
\end{figure*}

\begin{figure*}
	\begin{tabular}{ll}
		(a) & (b)\\
		\includegraphics[width=0.4\columnwidth]{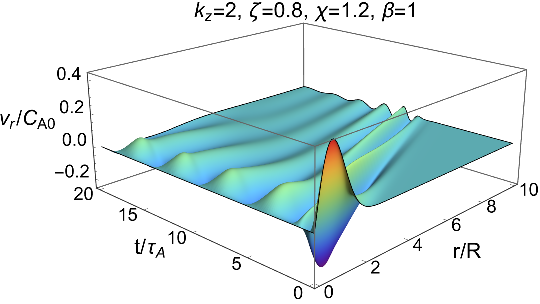}&\includegraphics[width=0.4\columnwidth]{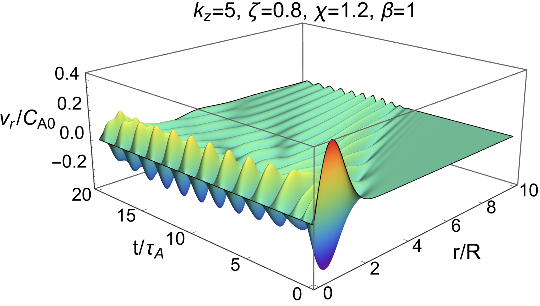}\\
		(c) & (d)\\
		\includegraphics[width=0.4\columnwidth]{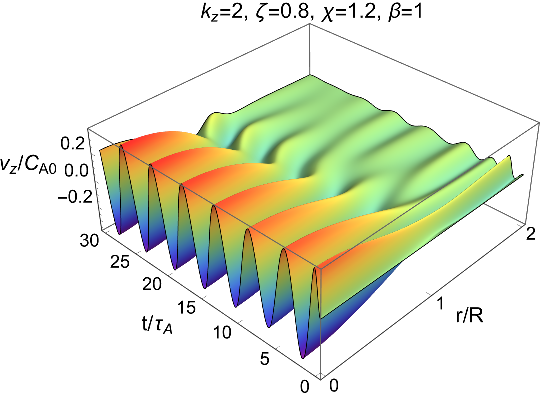}&\includegraphics[width=0.4\columnwidth]{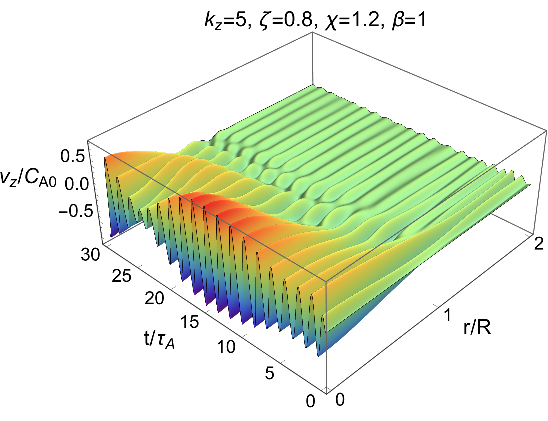}\\
		(e) & (f)\\
		\includegraphics[width=0.4\columnwidth]{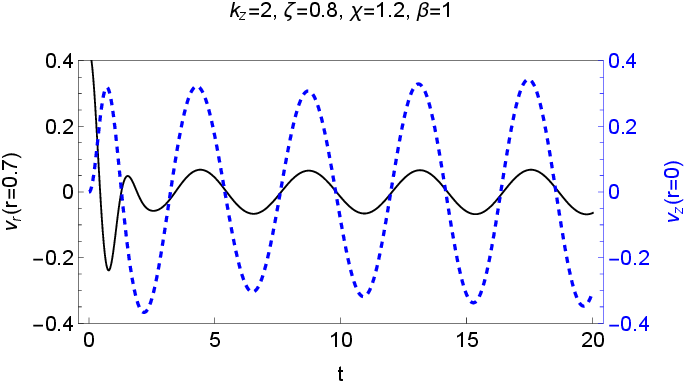}&\includegraphics[width=0.4\columnwidth]{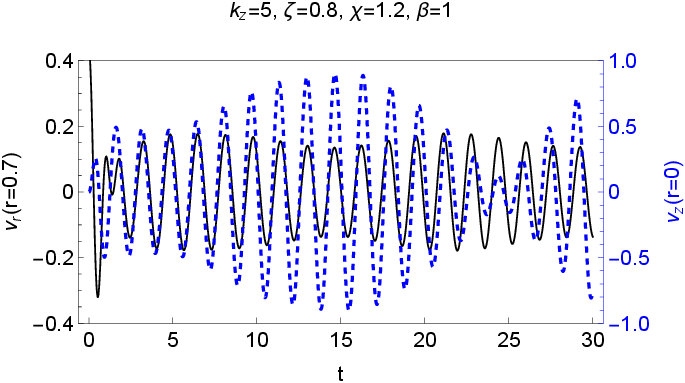}\\
		(g) & (h)\\
		\includegraphics[width=0.4\columnwidth]{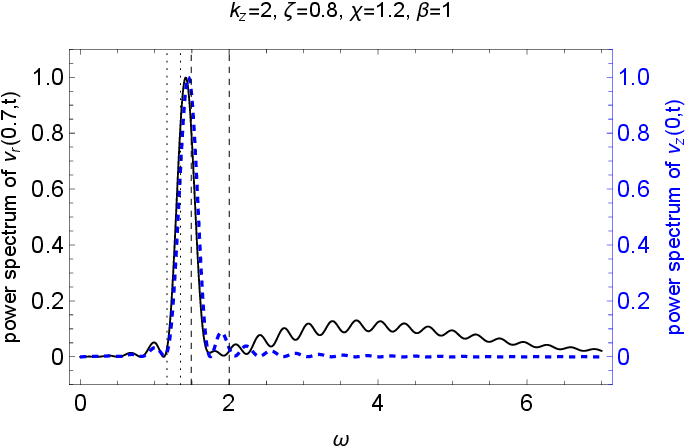}&\includegraphics[width=0.4\columnwidth]{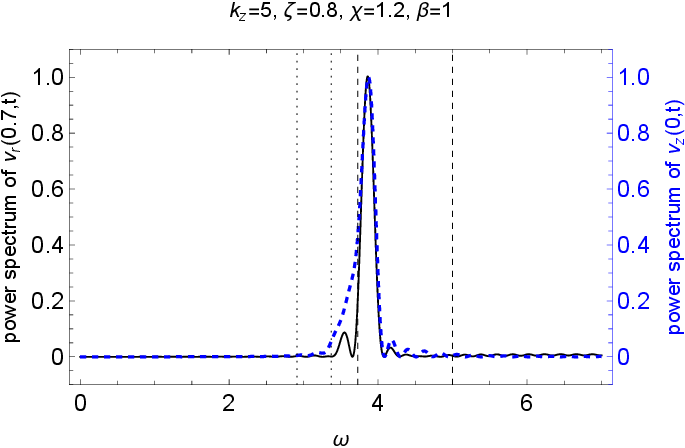}
	\end{tabular}
	\caption{\linespread{1}\selectfont Same as Fig. \ref{fig:beta02} but for $\beta_i=1$. Left and right panels illustrate results for $k_z=2$ and $k_z=5$, respectively.}
	\label{fig:beta1}
\end{figure*}

Figure \ref{fig:beta02} presents the results for ($\beta_i,\zeta,\chi$)=($0.2, 0.8, 1.2$) for which $T_{00}=0.49 T_{\infty}$. Here, we obtain the results for two wavenumbers $k_z=5$ and $k_z=20$. The first and second rows of the Figure display the spatio-temporal evolution of the radial and longitudinal velocity perturbations, respectively. The third row shows the temporal evolution of the radial velocity perturbation at $r=0.7$ (the location of the maximum value of the initial $v_r$ perturbation) and the longitudinal velocity perturbation at $r=0$, while the fourth row presents the corresponding power spectra derived from these data. As illustrated in panels (a) and (b) of the Figure, in both cases of $k_z=5$ and $k_z=20$ the sausage wave is leaky since the radial velocity perturbation decreases rapidly and leaks out of the tube in the radial direction.In panels (c) and (d) of the figure, the amplitude of the longitudinal velocity perturbation on the tube axis is almost one order of magnitude smaller than that of the initial radial velocity perturbation. The amplitude of the excited $v_z$ depends on the plasma-$\beta$ within the flux tube, with larger values of plasma-$\beta$ producing a larger longitudinal velocity perturbation \citep[see also][]{Ebrahimi2024}. The results further show that, in addition to the excitation of $v_z$ on the tube axis, the longitudinal velocity perturbation also grows in the exterior region around $r=2$, where it contributes to the leakage of the wave. This external $v_z$ perturbation should therefore not be identified as a slow sausage wave. In the present configuration, the slow sausage response is confined to the flux tube, whereas the external $v_z$ perturbation is associated with the outward-propagating disturbance responsible for wave leakage. This distinction is important because, unlike fast sausage waves, slow sausage waves are generally trapped in the standard cylindrical flux-tube model. In panel (g), the regions bounded by the vertical dashed and dotted lines correspond to the Alfv\'{e}n and slow continua, respectively. As depicted in panel (g) of Figure \ref{fig:beta02} (the blue dashed curve), the longitudinal velocity perturbation on the tube axis oscillates with a frequency in the slow continuum. Therefore, the excited longitudinal velocity perturbation, $v_z$, is identified as a slow sausage wave. Also, it is clear from panel (g) that the power spectrum of the radial oscillation consists of three characteristic peaks. The first peak that has a small power and located in the slow continuum corresponds with radial oscillation of the slow sausage wave. The second peak is in the Alfv\'{e}n continuum around $\omega_{A,e}$ (left vertical dashed line) which denote the excitation of a decay-less small amplitude fast sausage wave after which the initial fast wave leaks out of the tube. The third peak located out of the Alfv\'{e}n spectrum around $\omega =6$ corresponds with the leaky fast sausage wave. Frequency analysis, as illustrated in panel (h) for $k_z=20$, reveals that the radial velocity perturbation oscillates at a frequency outside the Alfv\'{e}n continuum, indicating a leaky fast sausage wave. At the same $k_z$, the longitudinal velocity perturbation on the tube axis oscillates predominantly at a frequency within the slow continuum, consistent with the excitation of a slow sausage wave. However, a weaker secondary peak is also present at the frequency corresponding to the maximum power of the radial perturbation, indicating that the longitudinal velocity perturbation also contains a small contribution associated with the leaky fast sausage wave.

As mentioned previously, for the step-function profile ($\alpha=\epsilon=\infty$), the cut-off wavenumber for $(\beta_i,\zeta,\chi)=(1,0.8,1.2)$ is $k_c^{(1)}=2.9$. Although the actual cut-off wavenumber may differ for the present model with continuous radial profiles, we adopt $k_c^{(1)}=2.9$ as an initial estimate and perform simulations for $k_z=2$ and $k_z=5$. For this parameter set, $T_{00}=0.96 T_{\infty}$, indicating that the pore is 4\% cooler than its surroundings. The results for $(\beta_i,\zeta,\chi)=(1,0.8,1.2)$ are presented in Figure \ref{fig:beta1}. It is clear from the figure that, for $k_z=2$ (left panels), the amplitude of the radial velocity perturbation decreases rapidly during the initial stage of the evolution and subsequently evolves into a much lower-amplitude, decay-less oscillation. During the initial stage, the wave exhibits characteristics of a leaky fast sausage wave. As depicted in panel (g) of Figure \ref{fig:beta1}, the broad frequency peak of $v_r$ around $\omega=4$ is associated with this initial stage. Thus, part of the wave energy leaks out of the flux tube, while another part of the perturbation energy contributes to the excitation of the longitudinal velocity perturbation. After this initial transient, the radial and longitudinal velocity perturbations oscillate predominantly at the same frequency, $\omega\simeq1.43$, which lies outside both the slow and Alfv\'{e}n continua. At this stage, as illustrated in panel (e) of Figure \ref{fig:beta1}, the dominant velocity perturbation is the longitudinal component. This trapped oscillation is therefore identified as a slow sausage mode. Since its frequency does not lie within the slow continuum, no slow resonant layer is present, and hence slow resonant absorption does not occur during the evolution. As illustrated in the right panels of Figure \ref{fig:beta1}, for $k_z=5$, the dominant frequency is $\omega\simeq3.8$. This frequency lies above the slow continuum between the internal and external Alfv\'{e}n speeds, indicating that the oscillation is a fast sausage wave. Although the longitudinal velocity perturbation has a relatively large amplitude, the substantial excitation of $v_z$ does not by itself imply a slow-mode character. Instead, it results from the stronger coupling between the radial and longitudinal perturbations associated with the relatively high value of $\beta=1$.

\begin{figure}
	\includegraphics[width=0.5\columnwidth]{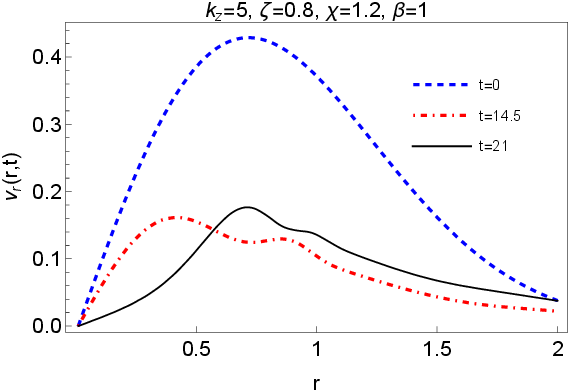}
	\caption{\linespread{1}\selectfont Radial velocity perturbation versus $r$ at three different times $t=0,~14.5$, and 21 at which $v_r$ reaches its peaks. Location and time are in units of the loop radius $R$ and the Alfv\'{e}n time $\tau_A$, respectively.}
	\label{fig:vrr}
\end{figure}
As depicted in panel (f) of Figure \ref{fig:beta1}, the radial perturbation undergoes a modulation in its amplitude. Interestingly, similar modulations could be seen in the sausage wave oscillations observed by \cite{Morton2011} and \cite{Keys2018} in photospheric pores. To further illustrate the amplitude modulation of $v_r$ for the trapped mode with $\beta=1$, Figure \ref{fig:vrr} shows the radial profiles of $v_r$ at three selected times corresponding to maxima in its temporal evolution. At $t=0$, $14.5$, and $21$, the maximum of $v_r$ occurs at $r=0.74$, $0.4$, and $0.7$, respectively. These profiles show that the location of the maximum radial velocity perturbation shifts across the flux tube as the wave evolves. This spatial redistribution is consistent with the amplitude modulation of $v_r(r=0.7,t)$ shown in panel (f) of Figure \ref{fig:beta1}.

\begin{figure*}
	\begin{tabular}{ll}
		(a) & (b)\\
		\includegraphics[width=0.4\columnwidth]{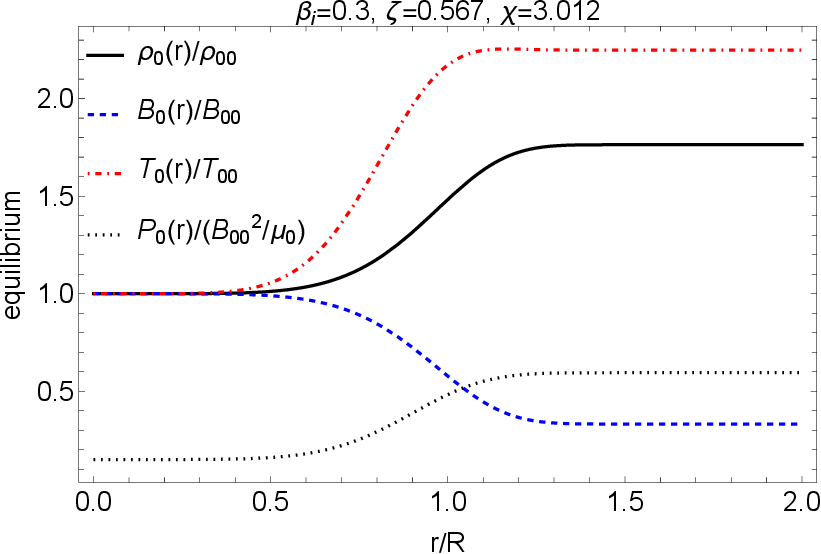}&\includegraphics[width=0.4\columnwidth]{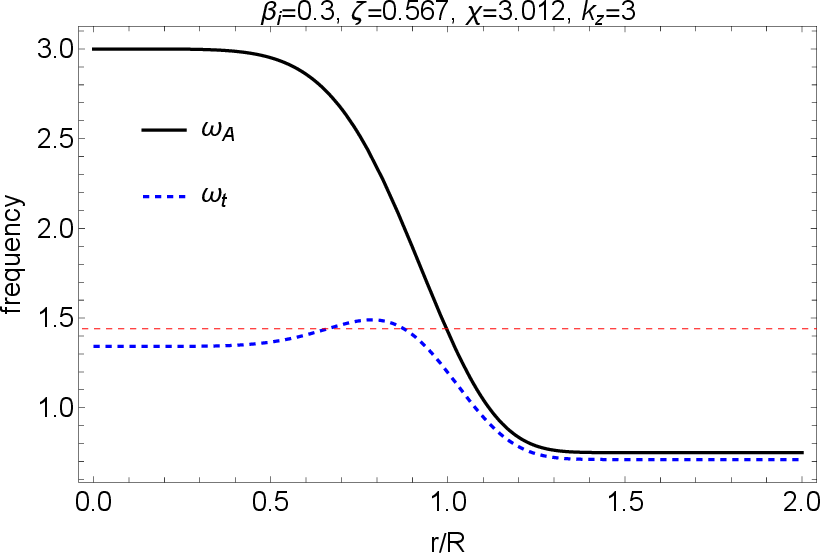}\\
		(c) & (d)\\
		\includegraphics[width=0.4\columnwidth]{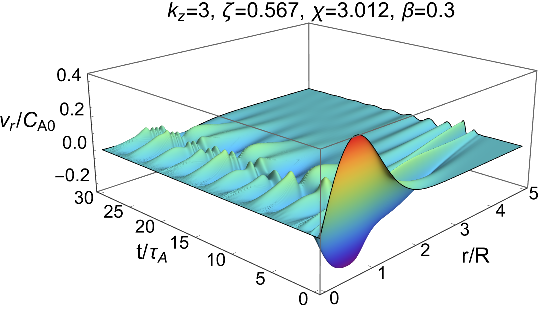}&\includegraphics[width=0.4\columnwidth]{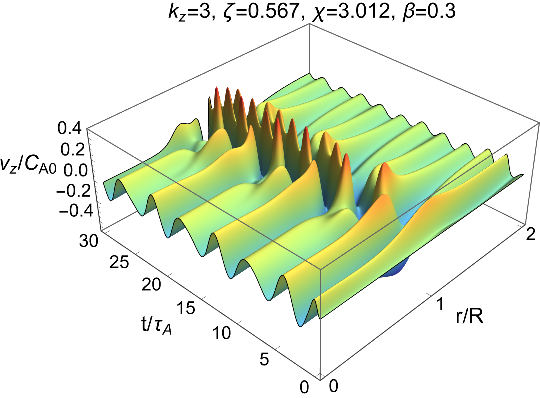}\\
		(e) & (f)\\
		\includegraphics[width=0.4\columnwidth]{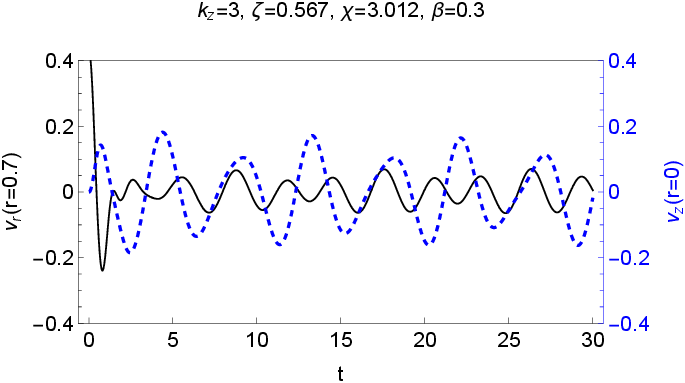}&\includegraphics[width=0.4\columnwidth]{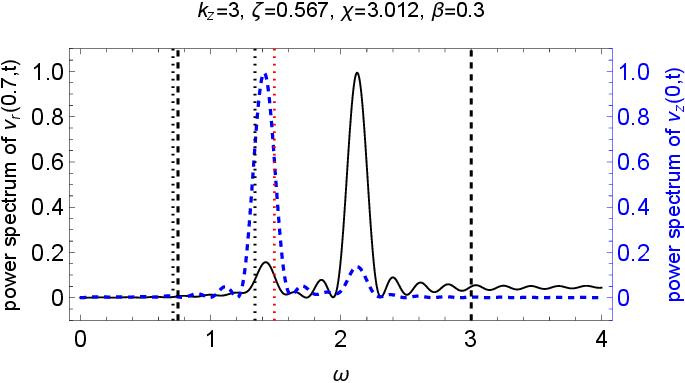}
	\end{tabular}
	\caption{\linespread{1}\selectfont Results for the parameter set $(\beta_i,\zeta,\chi)=(0.3,0.567,3.012)$ and $k_z=3$. Panel (a): radial profiles of the equilibrium quantities; Panel (b): background Alfv\'{e}n and slow frequencies; Panel (c): spatio-temporal behaviour of $v_r(r,t)$; Panel (d): spatio-temporal behaviour of $v_z(r,t)$; Panel (e): temporal evolution of $v_r(0.7,t)$ (solid line) and $v_z(0,t)$ (dashed line); Panel (f): power spectra of the time series presented in panel (e). The horizontal red dotted line in panel (b) denote the frequency associated with the left blue dashed peak in panel (f). The vertical red dotted line in panel (f) is the maximum frequency in the slow continuum plotted in panel (b). The definitions of the vertical dashed and dotted lines in panel (f) are the same as in Fig. \ref{fig:beta02}.}
	\label{fig:chen}
\end{figure*}

For the parameter set $(\beta_i,\zeta,\chi)=(0.3,0.567,3.012)$, the equilibrium temperature at the pore axis is $T_{00}=0.44T_{\infty}$, indicating that the pore is substantially cooler than its surroundings. Figure \ref{fig:chen}(a) shows the radial profiles of the equilibrium quantities, while panel (b) displays the Alfv\'{e}n and slow continua for $k_z=3$. Panels (c) and (d) illustrate the spatio-temporal evolution of the radial and longitudinal velocity perturbations, $v_r$ and $v_z$, respectively. The corresponding time series of $v_r(0.7,t)$ and $v_z(0,t)$ are presented in panel (e) as solid and dashed curves, respectively, while their power spectra are shown in panel (f). In the latter panel, the vertical black dotted lines indicate the internal and external tube frequencies, whereas the vertical black dashed lines mark the internal and external Alfv\'{e}n frequencies. As shown in panel (b), the slow continuum is non-monotonic, reaching its maximum frequency at $r=0.8$. The red dotted line in panel (f) marks this maximum frequency. Panels (c) and (e) show that the radial velocity perturbation undergoes a substantial reduction in amplitude within approximately half an oscillation period, after which it continues to oscillate with a smaller, modulated amplitude. The power spectra in panel (f) reveal two prominent frequency components, at approximately $\omega=2.1$ and $\omega=1.4$. The higher-frequency component is associated with the fast sausage wave and lies within the Alfv\'{e}n continuum, consistent with a trapped fast-wave response. The lower-frequency component is associated with the slow sausage wave and lies within both the slow and Alfv\'{e}n continua. In particular, the horizontal red dashed line in panel (b), corresponding to $\omega\simeq1.4$, intersects the slow continuum at $r\simeq0.66$ and $r\simeq0.88$, and the Alfv\'{e}n continuum at $r\simeq1$. These intersections identify two slow-resonance locations and one Alfv\'{e}n-resonance location for this frequency. The associated radial variation of the local continuum frequencies can produce phase mixing in the vicinity of these resonant layers. Consistent with this interpretation, panels (c) and (d) show pronounced radial structuring of both velocity components around $r\simeq1$. As the evolution proceeds, phase mixing generates increasingly fine spatial scales in the velocity perturbations, enhancing the effects of viscous dissipation. Consequently, viscosity can become dynamically significant at sufficiently advanced stages of the evolution, even for large Reynolds numbers, because the development of steep transverse gradients increases the efficiency of viscous dissipation \citep{Heyvaerts1983a}.

\subsection{The Effect of Viscosity}

Magnetic pore as a photospheric structure consists of a partially ionized plasma including neutrals. However, for the sake of simplicity we approximate it as a fully ionized plasma. In this case, the kinematic viscosity is obtained from the \cite{Spitzer1962} relation, namely, $\rho\nu=2.21\times 10^{-17} T^{5/2}/\ln \Lambda~ \rm{kg~m^{-1}~s^{-1}}$ in which, $T$ is temperature and $\ln \Lambda$($\approx 10$ for solar atmospheric conditions) is the Coulomb logarithm. Assuming typical values of the equilibrium quantities as $T\in(10^4 , 10^5)~\rm{K}$, $\rho\in (10^{-8} , 10^{-6}) ~ \rm{g~cm^{-3}}$, $B_0\in (100,1000) ~\rm{G}$ and $R=10^6 ~\rm{m}$, we obtain the kinematic viscosity in the range $\nu\in (2.21\times 10^{-5}, 0.7)~ \rm{m^2~s^{-1}}$ and the viscous Reynolds number in the range $R_v \in (4\times 10^{9},1.3\times 10^{14})$. 

Here, we investigate the effect of viscosity on the damping of sausage waves by examining the evolution of the wave energy within the flux tube. To derive the corresponding energy equation, we obtain the scalar products of Eqs. \eqref{mhd2}–\eqref{mhd4} by $\mathbf{v}_1$, $\mathbf{B}_1/\mu_0$, and $p_1/(\gamma P_0)$, respectively. Summing the resulting equations and doing some algebra yields
\begin{equation}\label{energy}
	\begin{split}
		&\frac{\partial}{\partial t}\mathcal{E}=-\nabla\cdot \mathbf{F}\\
		&-\frac{p_1}{\gamma P_0}\mathbf{v}_1\cdot\nabla P_0+\frac{1}{\mu_0}\mathbf{v}_1\cdot\left[(\nabla\times \mathbf{B}_0)\times \mathbf{B}_1\right]\\
		&+\rho_0\nu \mathbf{v}_1\cdot\left[\nabla^2\mathbf{v}_1+\frac{1}{3}\nabla(\nabla\cdot\mathbf{v}_1)\right],
	\end{split}
\end{equation}
where $\mathcal{E}$ and $\mathbf{F}$ denote the perturbation energy density and energy-flux density, respectively, given by
\begin{equation}\label{e_den}
	\mathcal{E}=\frac{1}{2}\rho_0 |\mathbf{v}_1|^2+\frac{|\mathbf{B_1}|^2}{2\mu_0}+\frac{p_1^2}{2\gamma P_0},
\end{equation}
\begin{equation}\label{f_den}
	\mathbf{F}=p_1\mathbf{v}_1+\frac{1}{\mu_0}\mathbf{B}_1\times(\mathbf{v}_1\times\mathbf{B}_0).
\end{equation}
In Eq. \eqref{e_den}, the three terms correspond to the kinetic, magnetic, and compressive contributions, respectively.
In Eq. \eqref{energy}, the second and third terms on the right hand side describe the exchange of energy between the wave and the non-uniform equilibrium pressure and magnetic field, respectively. The last term represents the effect of viscosity and is responsible for the viscous dissipation of wave energy.

By integrating Eq. \eqref{energy} over the computational volume and subsequently over time from $0$ to $t$, we obtain
\begin{equation}
	\begin{split}
		&E(t)-E(0)=-\int_0^t\oint_{\partial V}\mathbf{F}\cdot d\mathbf{S} dt'\\
		&-\int_0^t\int_V \left(\frac{p_1}{\gamma P_0}\mathbf{v}_1\cdot\nabla P_0-\frac{1}{\mu_0}\mathbf{v}_1\cdot\left[(\nabla\times\mathbf{B}0)\times\mathbf{B}_1\right] \right) dV dt'\\
		&+\int_0^t\int_V\rho_0\nu\mathbf{v}_1\cdot\left[\nabla^2\mathbf{v}_1+\frac{1}{3}\nabla(\nabla\cdot\mathbf{v}_1)\right]dV dt',
	\end{split}
\end{equation}
where
\begin{equation}
	E(t)=\int_V\mathcal{E} dV.
\end{equation}
The surface-integrated energy flux can be neglected because the wave does not reach the outer boundary at $r=D$ during the time interval considered. Also, the radial energy flux vanishes at the symmetry axis $r=0$. Hence, in the absence of viscosity ($\nu=0$), the conserved quantity that represents the energy content of the wave is
\begin{equation}
	\begin{split}
		&E_{\rm total}(t)=E(t)\\
		&+\int_0^t\int_V \mathbf{v}_1\cdot\left(\frac{p_1}{\gamma P_0}\nabla P_0-\frac{1}{\mu_0}\left[(\nabla\times\mathbf{B}_0)\times\mathbf{B}_1\right] \right) dV dt'.
	\end{split}
\end{equation}
where the second term on the right-hand side represents the cumulative work done by the effective pressure and magnetic force terms within the computational volume. For a flux tube with uniform equilibrium magnetic field and pressure, this term vanishes. Note that for the initial conditions \eqref{init} we have
\begin{equation}
	E_{\rm total}(t=0)=E(t=0)=\int_V\frac{1}{2}\rho_0 v_1^2(r,0) dV.
\end{equation}
\begin{figure}
	\begin{tabular}{ll}
		\includegraphics[width=0.5\columnwidth]{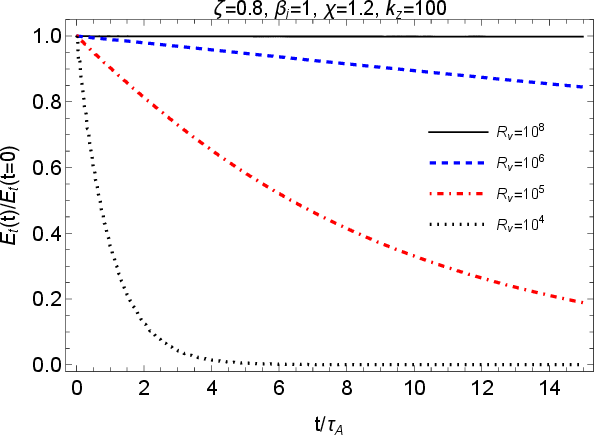}\\
		\includegraphics[width=0.5\columnwidth]{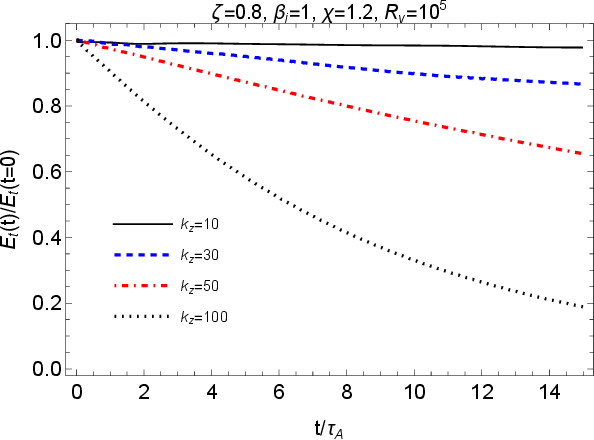}\\
		\includegraphics[width=0.5\columnwidth]{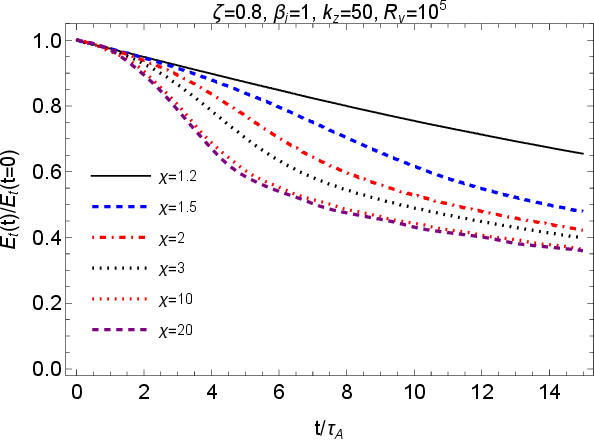}
	\end{tabular}
	\caption{\linespread{1}\selectfont Temporal evolution of the normalized total wave energy, $E_t(t)/E_t(t=0)$ for $\alpha=\epsilon=6$, $\zeta=0.8$ and $\beta_i=1$. Top: dependence on $R_v$ for $\chi=1.2$ and $k_z=100$. Middle: dependence on $k_z$ for $\chi=1.2$ and $R_v=10^5$. Bottom: dependence on the magnetic-field contrast $\chi$ for $k_z=50$ and $R_v=10^5$.}
	\label{fig:te}
\end{figure}

The energy evolution shown in Figure~\ref{fig:te} demonstrates the dependence of the wave-energy decay on the Reynolds number, longitudinal wavenumber, and magnetic-field contrast. In the upper panel, for $k_z=100$, the normalized total energy remains almost constant for $R_v=10^8$, indicating very weak viscous dissipation. As $R_v$ decreases, the energy decays progressively faster: for $R_v=10^6$, the energy decreases only moderately, whereas for $R_v=10^5$ it falls to about $20\%$ of its initial value by $t=15$. The strongest dissipation occurs for $R_v=10^4$, for which the wave energy is almost completely depleted by $t\simeq4$. Thus, the results clearly demonstrate the expected enhancement of viscous damping with decreasing Reynolds number.

The middle panel shows that, at a fixed $R_v=10^5$, increasing $k_z$ substantially enhances the energy decay. The cases with $k_z=10$ and $30$ retain about $98\%$ and $87\%$ of their initial energy, respectively, at $t=15$, whereas the energy decreases to approximately $65\%$ and $20\%$ for $k_z=50$ and $100$. This strong dependence on $k_z$ is consistent with the enhanced generation of small-scale spatial gradients at larger longitudinal wavenumbers, which increases viscous dissipation. The lower panel further shows that increasing the magnetic-field contrast, $\chi=B_{00}/B_\infty$, also enhances the energy decay. For $k_z=50$ and $R_v=10^5$, the energy remaining at $t=15$ decreases from about $0.66$ for $\chi=1.2$ to $0.36$ for $\chi=20$. The increasingly similar decay curves for $\chi=10$ and $20$ suggest that the effect begins to saturate at larger magnetic-field contrasts. Overall, the results indicate that viscous energy dissipation is enhanced by decreasing $R_v$, increasing $k_z$, and increasing the radial magnetic-field contrast.

It is worth noting that the feedback of viscous heating on the equilibrium is not included in the present model. This approximation is justified provided that the accumulated viscous heating remains sufficiently small over the timescale considered or is balanced by unresolved cooling processes.

\section{Summary and Conclusions} \label{conclusion}

In this study, we investigated the conditions for the existence of trapped sausage modes in photospheric magnetic pores. We modelled a magnetic pore as a straight cylindrical flux tube with a circular cross-section, containing a non-uniform magnetic field aligned with the tube axis and an underdense plasma whose density varies smoothly in the radial direction. Requiring the pore temperature to be lower than that of its environment places an upper bound on the plasma-$\beta$ value on the tube axis, with this bound depending on the equilibrium parameters. For a non‑zero plasma-$\beta$ and in the presence of viscosity, we derived the linearised equations of motion governing sausage‑mode perturbations.

We first examined the cut-off wavenumber, $k_c$, derived by \cite{Roberts1984} for a photospheric magnetic flux tube with step-function radial equilibrium profiles. We investigated the regions of the $(\beta_i,\zeta)$ parameter space in which the corresponding cut-off relation yields real or imaginary values. Our results reveal a narrow region under photospheric pore conditions in which $C_{t,i}^2<C_{A,e}^2<C_{s,i}^2$ and the cut-off wavenumber is real. Within this region, the real value of $k_c$ can provide an initial estimate of the cut-off wavenumber for the present model, in which the equilibrium quantities vary continuously in the radial direction. In two other regions of parameter space, characterised by $C_{A,e}^2<C_{t,i}^2<C_{s,i}^2$ and $C_{A,e}^2>C_{s,i}^2>C_{t,i}^2$, respectively, the cut-off relation yields an imaginary value. This imaginary value should not be interpreted as a physical cut-off wavenumber, rather, it indicates that the cut-off condition employed in the derivation of \cite{Roberts1984} cannot be satisfied by a real longitudinal wavenumber. Consequently, the relation does not provide a physical criterion for the transition between trapped and leaky waves in these parameter regimes.

To investigate the sausage wave dynamics in the radially structured model, we numerically solved an initial-value problem using the PDE interface of COMSOL Multiphysics\textregistered. We examined the spatio-temporal evolution of sausage-mode perturbations for three parameter sets, $(\beta_i,\zeta,\chi)=(0.2,0.8,1.2)$, $(1,0.8,1.2)$, and $(0.3,0.567,3.012)$. The results for $(\beta_i,\zeta,\chi)=(1,0.8,1.2)$ which has a real cut-off wavenumber based on \cite{Roberts1984} relation, demonstrate that the trapped or leaky character of the excited perturbations could be determined from the cut-off relation derived for a discontinuous equilibrium. Results for $(\beta_i,\zeta,\chi)=(0.3,0.567,3.012)$  show that trapped wave behaviour can occur even when that relation yields an imaginary cut-off wavenumber. For $(\beta_i,\zeta,\chi)=(0.2,0.8,1.2)$, for which the cut-off wavenumber is also imaginary, the excited fast sausage waves are leaky over the range of longitudinal wavenumbers investigated. These results demonstrate that the applicability of the cut-off relation derived for a discontinuous equilibrium depends on the equilibrium parameters and that the trapped or leaky character of fast sausage waves in radially structured magnetic pores must be assessed by solving the wave-evolution problem for the specific equilibrium.

Our analysis of viscous dissipation shows that the decay of sausage-wave energy depends strongly on the Reynolds number, longitudinal wavenumber, and magnetic-field contrast. The energy decay becomes more pronounced as the Reynolds number decreases, the longitudinal wavenumber increases, and the radial magnetic-field contrast becomes stronger. These trends are consistent with enhanced viscous dissipation associated with increasingly fine spatial gradients. In particular, phase mixing can progressively generate small radial scales, thereby enhancing viscous damping even when the Reynolds number is large. The results also suggest that the influence of magnetic-field contrast on energy dissipation becomes less pronounced at sufficiently large contrasts. This analysis highlights the importance of phase mixing in facilitating the dissipation of sausage-wave energy, while the quantitative damping rates remain dependent on the adopted transport coefficients and equilibrium parameters. The feedback of viscous heating on the equilibrium was neglected in the present model.

A limitation of the present model is the absence of longitudinal stratification of the equilibrium quantities. Realistic photospheric pores are vertically structured, with the magnetic-field strength, mass density, and gas pressure varying over characteristic scale heights that may be comparable to or smaller than the pore radius. Consequently, gravity and longitudinal stratification can introduce additional spatial scales and modify the propagation and mode properties of MHD waves. The present model should therefore be regarded as an idealized waveguide in which the equilibrium is structured only in the transverse direction. This approximation is nevertheless useful for isolating the effects of radial structuring on sausage-mode dynamics and for establishing a controlled reference model for wave trapping, leakage, and energy evolution. In particular, the results presented here characterize effects arising from the radial profiles of the magnetic field and density without the additional coupling and modification introduced by vertical stratification. As emphasized by \cite{Chen2021}, extending such models to include gravity and longitudinally varying equilibrium quantities is substantially more involved and introduces additional characteristic scales. The present results can therefore serve as a useful reference and benchmark for future studies incorporating vertical stratification, and may provide a useful basis for interpreting observed sausage-wave signatures in photospheric pores, particularly when assessing the role of transverse structuring in determining wave trapping, leakage, and energy transport.

Also, it is important to emphasise that the present analysis is restricted to the linear regime. Genuinely non-linear effects such as non-linear mode coupling and harmonic generation, non-linear steepening, amplitude-dependent periods and damping rates, and the non-linear evolution and saturation of instabilities are not captured within this framework and may modify the behaviour reported here. A comprehensive assessment of these processes requires high-resolution dissipative MHD simulations capable of resolving the full non-linear evolution.

\subsection*{Acknowledgements}
The author gratefully acknowledges the anonymous referee for their thoughtful comments and constructive suggestions, which significantly improved the clarity and quality of this manuscript.
\bibliographystyle{bib_style} 
\bibliography{refs}
\end{document}